\documentclass[prb,showpacs,raggedbottom,twocolumn,superscriptaddress,floatfix,amsmath]{revtex4}
\usepackage[latin1]{inputenc}
\usepackage[english]{babel}
\usepackage[T1]{fontenc}

\usepackage{graphicx}
\usepackage{amssymb}
\usepackage{amsmath}
\usepackage{amscd}
\usepackage{eucal}
\usepackage{color}
\usepackage{bm}

\newcommand{\rmd}{{\rm d}}

\newcommand{\half}{{\textstyle{\frac{1}{2}}}}

\newcommand{\eps}{\epsilon}

\newcommand{\eminus}{ e^{\operatorname{-}} }

\newcommand{\kB}{k_{\rm B}}
\newcommand{\Kn}{K_{\rm n}}

\newcommand{\tildei}{\tilde{\imath}}
\newcommand{\tildej}{\tilde{\jmath}}
\newcommand{\tilden}{\widetilde{n}}
\newcommand{\tildem}{\widetilde{m}}

\definecolor{DarkGreen}{rgb}{0,0.7,0}

\begin{document}

\title{
Thermodynamic and quantum bounds on nonlinear DC thermoelectric transport}

\author{Robert S.~Whitney}
\affiliation{
Laboratoire de Physique et Mod\'elisation des Milieux Condens\'es (UMR 5493), 
Universit\'e Grenoble 1 and CNRS, Maison des Magist\`eres, BP 166, 38042 Grenoble, France.}

\date{\today}
\begin{abstract}
I consider the non-equilibrium DC transport of electrons through a quantum system with a thermoelectric response. 
This system may be any nanostructure or molecule modeled by the nonlinear scattering theory which 
 includes Hartree-like electrostatic interactions exactly, and certain dynamic interaction effects (decoherence and relaxation) phenomenologically. 
This theory is believed to be a reasonable model when single-electron charging effects are negligible.  
I derive three fundamental bounds for such quantum systems coupled to multiple macroscopic reservoirs, one of which may be superconducting.  These bounds affect nonlinear heating (such as Joule heating), work and entropy production.  Two bounds correspond to the first law and second law of thermodynamics in classical physics.  The third bound is {\it quantum} (wavelength dependent), and is as important as the thermodynamic ones in limiting the capabilities of mesoscopic heat-engines and refrigerators.  The quantum bound also leads to Nernst's unattainability principle that the quantum system cannot cool a reservoir to absolute zero in a finite time, although it can get exponentially close.  
\end{abstract}



\maketitle

\begin{figure}[b]
\includegraphics[width=\columnwidth]{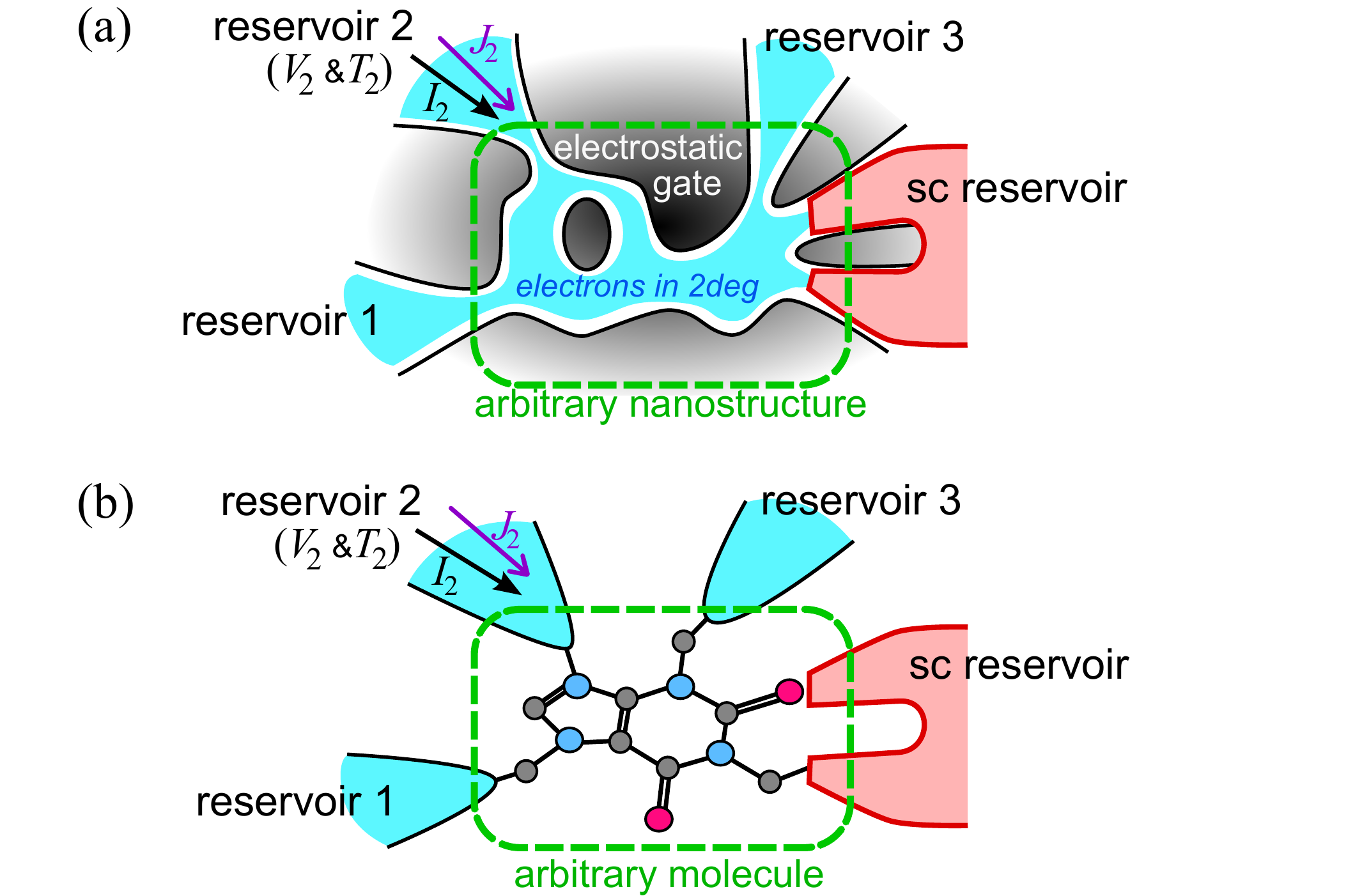}
\caption{\label{Fig:circuit}
Quantum systems with $\Kn=3$ normal reservoirs and
one superconducting (sc) reservoir. 
Examples include (a) nanostructures such as those defined by electrostatic gates
on top of a two-dimensional electron gas (2deg), or (b) molecules. 
Charge and heat current, $I_i$ and  $J_i$, are positive if they flow from reservoir $i$
into the quantum system.
}
\end{figure}

\section{Introduction}
In classical physics, currents in resistive circuits are dissipative;
they emit Joule heat that increases the entropy of the environment of the circuit.  
Devices with thermoelectric responses \cite{books,DiSalvo-review,Shakouri-reviews}
can be used as 
heat-engines (using heat flows to create electrical power) 
or heat pumps (using electrical power to create heat flows).
These are also dissipative, since they increase the entropy
of the environment  (at least a little bit) during their operation.
The laws of thermodynamics give us strict 
bounds on the capabilities of such thermoelectric devices.

There is currently great interest in using quantum systems as  thermoelectric heat-engines or refrigerator,
particularly nanostructures  \cite{Pekola-reviews,Zebarjadi2007,
Wierzbicki2011,Gunst2011,Rajput2011,Glatz2011,SSJB2012,
Sothmann-Buttiker2012,Trocha2012,2012w-pointcont}
and molecules \cite{Casati2008,Nozaki2010,
Entin-Wohlman2011,Saha2011,Karlstrom2011,Liu2011}.
However quantum responses are often non-local, 
since thermal equilibration occurs on lengthscales 
very much bigger than the size of the system, with equilibrium only being
established deep in the reservoirs coupled to the system. 
Classical models of transport account neither for this non-locality, nor for the wave-nature of quantum particles.
It is natural to ask what the bounds are for these quantum 
devices, and how they compare to the bounds on classical
devices.  There are many works
asking whether quantum systems are bounded by the laws of thermodynamics
\cite{QuantumThermodyn-book,
Wheeler1991,Abe2003,Brandao2008,Sagawa2008,Jacquet2009,delRio2011,Bruneau2012},
particularly for systems modeled by Markovian or Bloch-Redfield master equations,
see Refs.~[\onlinecite{QuantumThermodyn-book,Maruyama2009,Levy2012,Kolar2012}] and references therein. 
However the fundamental bounds on {\it thermoelectric} quantum systems remains an 
open question \cite{Bruneau2012}.
Here, I consider the fully nonlinear response 
of a broad class of thermoelectric quantum systems (see Fig.~\ref{Fig:circuit}), and show that they
obey the bounds imposed by the laws of thermodynamics,
while also suffering a quantum bound on their behavior.

The Landauer-B\"uttiker scattering theory 
\cite{review-Blanter-Buttiker}
is a widely used theory for modeling
such electron flows through quantum systems; 
giving DC and AC conductances
\cite{Buttiker1993-96,Petitjean2009}, 
and nonlinear effects
\cite{Christen-ButtikerEPL96,Sanchez-Buttiker,Meair-Jacquod2012,Sanchez-Lopez2012}.  
It has equally been used to get linear thermal and thermoelectric effects 
\cite{
Engquist-Anderson1981,Sivan-Imry1986,Butcher1990,Claughton-Lambert,jw-epl,jwmb}.
However it is a delicate question whether it captures
dissipation and the associated laws of thermodynamics
(or their quantum equivalent). 
The standard answer is that the theory accurately models {\it reversible} scattering processes, 
while assuming that equilibration  occurs (irreversibly) in the reservoirs via 
dissipative processes which the theory does not detail. 
Hence, one might suspect that the theory would not 
capture how dissipation produces entropy.
This suspicion is wrong, the nonlinear scattering theory 
 is sufficient to give information about entropy production, despite the simplistic way in which it treats the
 physics in the reservoirs.
Indeed, Ref.~[\onlinecite{Bruneau2012}] 
showed the second law of thermodynamics to be a 
remarkably direct outcome of a nonlinear ``DC'' scattering theory 
\cite{footnote:DCresponse} 
for a quantum system of {\it non-interacting} electrons 
between two reservoirs  \cite{footnote:Jacquet}.  
Energy conservation leads to  
the systems also obeying the first law.

Here I show that the proof in Ref.~[\onlinecite{Bruneau2012}] works equally well 
for a non-equilibrium scattering theory 
that includes
Hartree-like {\it electron-electron interactions} 
(electrostatic charging effects) in a self-consistent 
and gauge-invariant manner.
This theory is the analogue for heat-currents \cite{2012w-pointcont}
of the nonlinear scattering theory for charge-currents 
\cite{Christen-ButtikerEPL96,Sanchez-Buttiker,Meair-Jacquod2012,Sanchez-Lopez2012};
although heat-currents are not conserved when charge-currents are. 
Such electrostatic interactions cannot be ignored for the nonlinear response
\cite{Christen-ButtikerEPL96,Sanchez-Buttiker,Meair-Jacquod2012,Sanchez-Lopez2012},
and entropy production is a purely nonlinear effect (vanishing in the linear-response regime).
By keeping only electrostatic interaction effects, this approach is by no means exact,
however it is a reasonable approximation for any system 
where the single-electron charging energy is small
compared to temperature or broadening of the quantum system's levels due to the coupling to the reservoirs
\cite{Christen-ButtikerEPL96}.
Within this approach, I prove that the DC response of any quantum system coupled to an 
arbitrary number of reservoirs --- one of which may be superconducting --- 
obeys the laws of thermodynamics.

The extension from two\cite{Bruneau2012} to many reservoirs requires no new physics,
but is complicated by the fact one can no longer use
the two-terminal Onsager relations for transmission matrices.  
However it extends our understanding of thermodynamics in quantum-transport in two 
ways.
\begin{itemize}
\item[(1)] Using the generalization to many reservoirs, 
one can include dynamic interaction effects (decoherence and relaxation) phenomenologically as additional fictitious reservoirs  
\cite{voltage-probes,deJong-BeenakkerJacquet2009,Jacquet2012}. 
Performing such an analysis,
I conclude that the classical laws of thermodynamics hold for partially coherent 
(as well as fully coherent) quantum-transport.

\item[(2)] 
A superconducting (sc) reservoir has no analogue in classical mechanics,
because the condensate is a macroscopic quantum state. 
In addition, charge-current can flow into (or out of) this condensate, 
but heat-current cannot. Thus a charge-current can pass into the sc reservoir, 
$I_{\rm sc}$ without any change in that reservoir's entropy; 
$\dot S_{\rm sc} = 0$.  
Despite this, I show that the laws of thermodynamics hold in the presence of such a reservoir.
\end{itemize}

There is also a {\it quantum bound} (qb) on the heat extracted via thermoelectric cooling, identical to Pendry's bound \cite{Pendry1983} for passive cooling
of a hot reservoir by a colder one (a fermionic
analogue\cite{footnote:universalSBlaw}  of the Stefan-Boltzmann law of 
black-body radiation).
No thermoelectric device, supplied with an arbitrary amount of DC power, can ever extract heat from 
a reservoir faster than the bound $J^{\rm qb} \propto N \, T^2$,
where $T$ is the reservoir temperature, and the contact between the device and the reservoir has  
$N$ transverse modes. 
This bound is {\it quantum}; it is only relevant when  
$N^{-1} \sim \lambda_{\rm F}^2/A_i\neq 0$,
for electron wavelength 
$\lambda_{\rm F}$, and contact cross-section $A_i$.

For thermoelectric refrigerators, the quantum and thermodynamic 
bounds {\it compete}, and the quantum bound can be a stronger constraint than that given by the thermodynamic bounds (the Carnot limit).
In contrast, the bounds {\it combine}
to give an upper limit on the power that a thermoelectric heat-engine can produce.

I find that the quantum bound means that the system
obeys Nernst's {\it unattainability principle}
(one of the two statements forming the third law of thermodynamics),
which states that a reservoir cannot be refrigerated to absolute zero in a finite time.
See for example Refs.~[\onlinecite{Wheeler1991,Levy2012,Kolar2012}]
for physical and Refs.~[\onlinecite{Belgiorno2003,Landsberg1956}]
for mathematical overviews.  
The quantum bound means that the systems considered here
have (at best) the critical behavior; 
i.e.~the reservoir {\it cannot} achieve zero temperature in a finite time,
but it can get exponentially close.
This rules out such systems as candidates for 
violating the unattainability principle in the manner 
Ref.~[\onlinecite{Kolar2012}] proposes.

\section{Explicit form of the bounds}
In this section, I give the three bounds,
before deriving them later in the article. 
I consider an arbitrary quantum system coupled to any number of
reservoirs, one of which may be superconducting.
Normal reservoirs are treated as free-electron gases
in thermal equilibrium, while a superconducting reservoir is treated as 
a condensate of Cooper pairs \cite{footnote:finite-SC-gap}.  
The heat-current flowing out of reservoir $i$ is $J_i$. 
The electrical power flowing out of reservoir $i$ is $P_i=V_iI_i$, 
with $V_i$ being the bias on the reservoir, and  $I_i$ being the charge-current flowing out of it. 
The rate of change of entropy in the $i$th reservoir is defined as 
${\dot S}_i = -J_i/T_i$ for reservoir 
temperature $T_i$. 

I define ${\dot E}_{\rm total}$ and ${\dot S}_{\rm total}$ as the rate of change of the total energy and entropy of the
quantum system and all reservoirs averaged over long times.
Under DC drive \cite{footnote:DCresponse},
the quantum system remains in the same steady-state on average 
over long times, so one can neglect its contribution to  ${\dot E}_{\rm total}$ and ${\dot S}_{\rm total}$.
Then scattering theory gives the thermodynamic laws,
\begin{eqnarray}
\dot{E}_{\rm total} \  =\ \ \sum_i (J_i+P_i) \ =\ 0 & & \hbox{ [$1^{\rm st}$ law]}
\label{Eq:1stlaw}
\\
\dot{S}_{\rm total} \  =\  \ \ \ \sum_i \ -J_i\big/ T_i \ \ \geq \ 0 & & \hbox{ [$2^{\rm nd}$ law]}
\label{Eq:2ndlaw}
\end{eqnarray}
where the sum is over all reservoirs.
The quantum bound (qb) on the heat current out of reservoir $i$ is
\begin{eqnarray}
J_i \leq J^{\rm qb}_i \equiv {\pi^2 \over 6 h} N_i \, (\kB T_i)^2  & & \hbox{ [quantum]}
\label{Eq:quantumbound}
\end{eqnarray}
where the quantum system couples to reservoir $i$ through $N_i$ transverse modes, and this equation defines 
$J^{\rm qb}_i$.  
Thus the rate of change of entropy in the $i$th reservoir is $\dot S_i \geq - J^{\rm qb}_i/T_i$.
Sections \ref{Sect:heat-engine} and \ref{Sect:refrigerator}
discuss how these bounds act as limits on heat-engines and refrigerators.

\section{Nonlinear scattering theory for heat transport}
The nonlinear Landauer-B\"uttiker scattering theory, developed for 
AC \cite{Buttiker1993-96,Petitjean2009}
and nonlinear-DC
\cite{Christen-ButtikerEPL96,Sanchez-Buttiker} charge flow, was recently extended 
to thermoelectric effects \cite{2012w-pointcont,Meair-Jacquod2012,Sanchez-Lopez2012}.
It includes (in a gauge-invariant and self-consistent manner)
Hartree-type electron-electron interaction effects,
which act as electrostatic charging effects.
Here, as in Ref.~[\onlinecite{2012w-pointcont}], I apply it to nonlinear heat flows
\cite{new-preprints,Meair-Jacquod-preprint,Lopez-Sanchez-preprint}. 

One starts with the
scattering matrix \cite{Christen-ButtikerEPL96,Sanchez-Buttiker,Meair-Jacquod2012,Sanchez-Lopez2012} 
${\cal S}_{ij}^{\mu \nu}$, for a $\nu$-particle  
entering the quantum system from reservoir $j$ to a $\mu$-particle leaving via reservoir $i$,
where $\mu,\nu$ are $\pm1$, with $+1$ for electrons with charge $\eminus$ or $-1$ for holes with charge $-\eminus$.
Here, an electron with energy $\eps$ is an {\it occupied} conduction-band state at energy 
$\eps+\eminus V_{\rm sc}$, while a 
hole at energy $\eps$ is an  {\it empty} conduction-band state at energy $-\eps+\eminus V_{\rm sc}$,
where $\eminus V_{\rm sc}$ is the superconductor's chemical-potential. 
These ``holes''  are not in a semiconductor's valence band, unlike those in Ref.~[\onlinecite{2012w-pointcont}]. 
Let the lead to reservoir $i$ have $N_i$ modes for all $\eps$.
The number of {\it open} modes in this lead can still be $\eps$-dependent; since a given mode 
can be closed off at a certain $\eps$ (purely reflected in 
the scattering matrix at $\eps$). 
If a superconducting (sc) reservoir is present, 
it is modeled by Andreev reflection of electron to hole  and vice versa.
Then, one takes $\eps=0$ at the sc reservoir's chemical-potential $\eminus V_{\rm sc}$;
(otherwise one can choose $\eps=0$ however is convenient).
In the nonlinear regime,
${\cal S}_{ij}^{\mu \nu}\big( \eps,{\bf v}_{\rm env} \big)$ must include the charging effects self-consistently, 
because ${\cal S}_{ij}^{\mu \nu}$ is a function of the charge-distribution in the quantum system, which is, in turn,
a function of ${\cal S}_{ij}^{\mu \nu}$ as well ${\bf v}_{\rm env}$
\cite{Christen-ButtikerEPL96,Sanchez-Lopez2012},
where ${\bf v}_{\rm env}$ is the set of all voltages and temperatures, $\{V_k,T_k\}$,
in the system's environment (reservoirs and gates).

All charge and heat transport properties
are then given by the transmission matrix,
$
{\cal T}_{ij}^{\mu \nu} \big(\eps, {\bf v_{\rm env}} \big)
= {\rm Tr} \left[\big[{\cal S}_{ij}^{\mu \nu}\big(  \eps, {\bf v_{\rm env}} \big) \big]^\dagger \ 
{\cal S}_{ij}^{\mu \nu}\big( \eps, {\bf v_{\rm env}} \big) \right] 
$
where the trace is over all transverse modes of the leads connecting the quantum system to the $i$ and $j$ reservoirs.
Any self-consistent analysis must of course respect gauge-invariance; 
i.e.~shifting all voltages by the same amount is tantamount to redefining the zero of energy,
and thus cannot change the physics.  
I ensure this by taking $\eps$ and all reservoir and gate voltages (and any screening potentials that
depend on them,
relative to the superconductor's chemical-potential,\cite{footnote:gauge-inv} 
$\eminus V_{\rm sc}$.
If there is no superconductor, one can take these relative to another reservoir voltage. 
In everything that follows, I will assume that 
${\cal T}_{ij}^{\mu\nu}$ has this gauge-invariance.

The results presented in this article are all based on only two properties of ${\cal T}_{ij}^{\mu \nu}$.  
The first property is due to the fact that
${\cal T}_{ij}^{\mu \nu}$ is the sum of the modulus squared 
of the elements of ${\cal S}_{ij}^{\mu \nu}$. This property is that  
for all $i,j,\mu,\nu, \eps,{\bf v}_{\rm env}$,
\begin{eqnarray}
{\cal T}_{ij}^{\mu \nu} \big(\eps,{\bf v_{\rm env}} \big)
\geq 0. 
\label{Eq:positivity-of-T}
\end{eqnarray} 
The second property is due to the unitary of ${\cal S}_{ij}^{\mu \nu}$ (particle conservation),
and is that for all $\eps,{\bf v}_{\rm env}$,  
\begin{eqnarray}
0&=& 
\sum_{j=1}^{\Kn} \sum_\nu \big[N_i\delta_{ij}^{\mu\nu} - {\cal T}_{ij}^{\mu \nu} \big( \eps,{\bf v}_{\rm env} \big) \big] 
 \nonumber \\
 &=&
\sum_{i=1}^{\Kn} \sum_{\mu} \big[N_i\delta_{ij}^{\mu\nu}-{\cal T}_{ij}^{\mu \nu} \big( \eps,{\bf v}_{\rm env} \big) \big] , 
\label{Eq:transmission-sum}
\end{eqnarray}
the $i,j$ sums are over the $\Kn$ normal reservoirs, while the $\mu,\nu$ sums are over $\pm1$ for electrons or holes. 

As Eqs.~(\ref{Eq:positivity-of-T},\ref{Eq:transmission-sum}) are the central results needed for all derivations in this work, let me emphasis that 
they apply for the Hartree approximation without further approximations.   
One could self-consistently solve for 
infinitely many local time-independent Hartree potentials (on a grid of vanishing cell size).  
The resulting (very complicated) self-consistent scattering matrix 
would be unitary, and so satisfy Eqs.~(\ref{Eq:positivity-of-T},\ref{Eq:transmission-sum}).

The charge current $I_i$ is the difference between the charge $\mu \eminus$ leaving and entering reservoir $i$. 
The heat current $J_i$ is the difference between the excess-energy $(\eps-\mu\eminus V_i)$ leaving and entering reservoir $i$ \cite{Butcher1990}.  Thus
\begin{eqnarray}
I_i \! &=& \! \sum_{j\mu\nu} \int_0^\infty {{\rm d}\eps \over h} \  \mu \eminus
\, \big[N_i \delta_{ij}^{\mu \nu}-{\cal T}_{ij}^{\mu \nu}(\eps) \big] \,  f_j^\nu (\eps),  
\label{Eq:I-initial}
\\
J_i \! &=& \! \sum_{j\mu\nu} \int_0^\infty {{\rm d}\eps \over h}\, (\eps\!-\!\mu \eminus V_i)
\big[N_i \delta_{ij}^{\mu \nu}-{\cal T}_{ij}^{\mu \nu}(\eps) \big]  \, f_j^\nu (\eps),\ \ 
\label{Eq:J-initial}
\end{eqnarray}
where the $j$-sum is over the $\Kn$ normal reservoirs, while 
$\mu\nu$-sums are over electrons ($+1$) and  holes ($-1$).
Here $\delta_{ij}^{\mu \nu}$ indicates the pair of Kronecker  $\delta$-functions,
$\delta_{ij}\delta_{\mu\nu}$.
The Fermi function for $\nu$-particles entering
from reservoir $j$ is
\begin{eqnarray}
f_j^\nu(\eps) = \left(1+\exp\left[(\eps - \nu \eminus V_j)\big/ (\kB T_j) \right] \right)^{-1}.
\label{Eq:f}
\end{eqnarray} 
Comparing Eqs.~(\ref{Eq:I-initial}) and (\ref{Eq:J-initial}), one can easily see that
\begin{eqnarray}
J_i = -V_i \, I_i + \sum_{j\mu\nu} \int_0^\infty {{\rm d}\eps \over h}\, \eps\, 
\,  \big[N_i\delta_{ij}^{\mu\nu}-{\cal T}_{ij}^{\mu \nu}(\eps) \big] f_j^\nu(\eps),
\label{Eq:J-initial-2}
\end{eqnarray}
If a superconductor is present, then the charge-current into it 
(in the form of Cooper pairs), $I_{\rm sc}$, is given by Kirchoff's law of 
current conservation. 
In contrast, heat current is not conserved. 
However, the heat  flow through a boundary at the surface of 
a superconducting reservoir must be zero,  
since an electron at energy $\eps$
is Andreev reflected as a hole with the same energy.
Thus  
\begin{eqnarray}
I_{\rm sc} = - \sum_j I_j,  \qquad  
J_{\rm sc} =0,
\label{Eq:Isc+Jsc}
\end{eqnarray}
where again the $j$-sum is over the $\Kn$ normal reservoirs.

The total heat absorbed by the quantum system, $J_{\rm total}$, is the sum of  Eq.~(\ref{Eq:J-initial-2}) 
over all reservoirs. Then $J_{\rm total} = -\sum_{i=1}^{\Kn} V_i \, I_i $, since  Eq.~(\ref{Eq:transmission-sum})
means that the second term in Eq.~(\ref{Eq:J-initial-2})  cancels. 
Thus far, voltages were relative to the sc reservoir, if we take them relative to another reference,
\begin{eqnarray}
J_{\rm total} = -V_{\rm sc}\, I_{\rm sc} - \sum_{i=1}^{\Kn} V_i \, I_i .
\label{Eq:Joule-total}
\end{eqnarray}
Kirchoff's law in Eq.~(\ref{Eq:Isc+Jsc}) means that changing all voltages by the same amount
does not change the total heat flow, $J_{\rm total}$.
Thus $J_{\rm total}$ respects gauge-invariance whenever $I_i$ does, where 
one recalls that $I_i$ is a self-consistent function of all lead and gate voltages and temperartures \cite{Sanchez-Lopez2012}.
Whenever $J_{\rm total}$ is non-zero, heat-currents are not conserved in the quantum system;
the quantum system is a heat-sink for $J_{\rm total}>0$, 
or a heat-source for $J_{\rm total}<0$.

\section{Zeroth law of thermodynamics}

The definition of equilibrium is one 
of the various statements that together form the zeroth law.
It is defined as a state for which one can cut the system 
into parts in any manner, and one will find {\it no} charge and 
heat flow between those parts.
Thus, if a system in equilibrium consists of multiple reservoirs (evidently in equilibrium with each other) all coupled to each other through a 
quantum system,
then there can not be any charge or heat current through that quantum system.
It is trivial to show that any quantum system, placed between an arbitrary number of reservoirs all in equilibrium with each other, must obey this consequence of the zeroth law; 
the scattering matrix ${\cal S}$ is unitary, this means that the 
transmission matrix obeys Eq.~(\ref{Eq:transmission-sum}).
Substituting this into Eqs.~(\ref{Eq:I-initial},\ref{Eq:J-initial}), 
one can see that charge and heat currents
are zero whenever $T_j=T_0$ and $V_j=V_0$ for all $j$.

\section{First law of thermodynamics and Joule heating}
Eq.~(\ref{Eq:Joule-total}) has the form of a {\it classical} Joule heating term;
i.e.\ {\it voltage $\times$ current}. 
This is due to the fact that
energy conservation ensures that the heat emitted
(or absorbed) is equal and opposite to electrical power supplied (or generated).
Thus, one arrives directly at the first law of thermodynamics, Eq.~(\ref{Eq:1stlaw}).
However it is important to note that this is only true upon summing over all reservoirs,
there is no particular relation between the power flowing out of reservoir $i$, 
$P_i=V_iI_i$,
and the heat-current out of that reservoir, given in Eq.~(\ref{Eq:J-initial-2}).
 
As a first example, consider a quantum system with two normal reservoirs at the same temperature but different biases, Kirchoff's law means that $I_2=-I_1$, and hence
$P_{\rm total}= (V_1-V_2)I_1$. Here $P_{\rm total}>0$, so the reservoirs supply power to the quantum system, the total heat current into the quantum system, 
$J_{\rm total}=-P_{\rm total} <0$, which corresponds to Joule heating.
If reservoir 1 is superconducting, then all this heat flows into reservoir 2.
Thus superconductors filter out the heat flow generated by Joule heating 
(along with any other heat flow).

\begin{figure}
\includegraphics[width=\columnwidth]{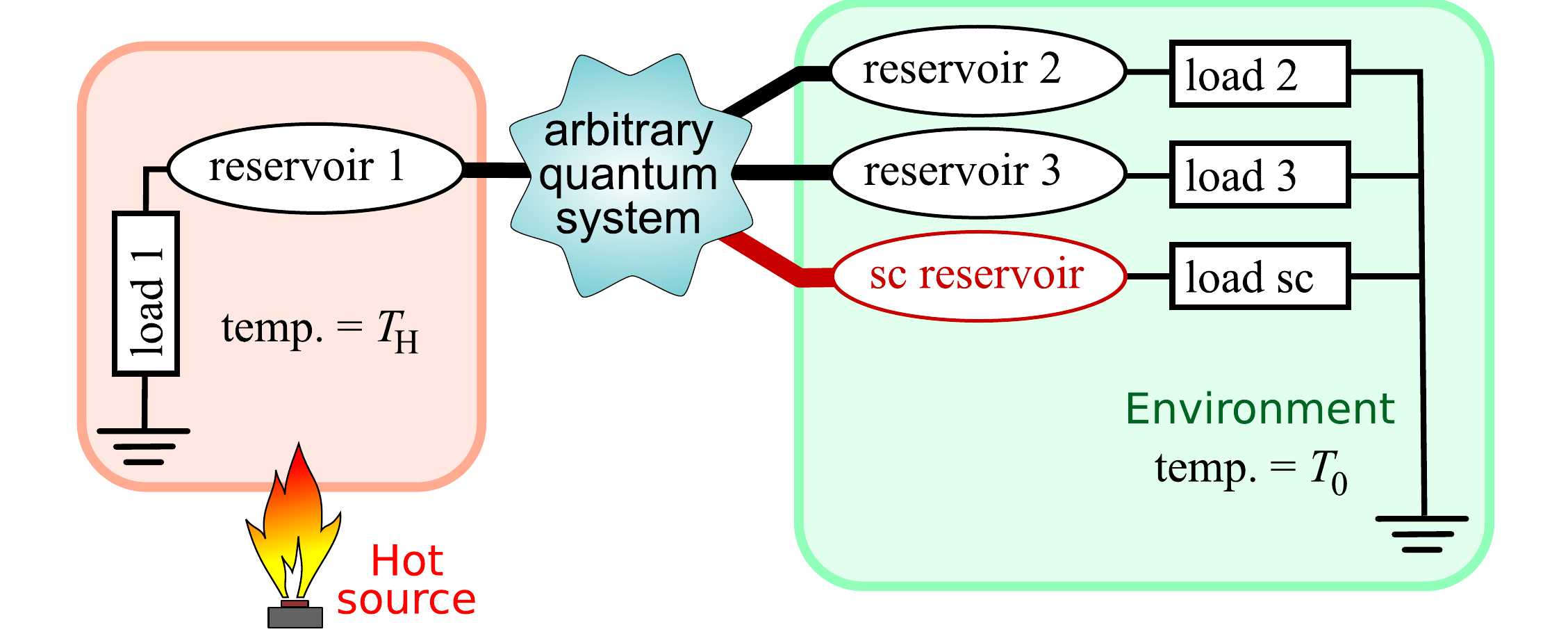}
\caption{\label{Fig:heatengine}
A quantum system with a finite thermoelectric response being used as a heat-engine,
providing power to a set of loads. 
The sc reservoir is superconducting, 
but ``load sc'' is not (its simply the load coupled to the sc reservoir).
}
\end{figure}

As a second example, the quantum system in Fig.~\ref{Fig:heatengine}
is assumed to have a finite thermoelectric response.  If one reservoir is heated, the quantum system generates
a total electrical power of $P_{\rm gen} =-P_{\rm total}= -\sum_i V_i I_i$ at the loads. 
Here $V_i$ and $I_i$ have opposite signs, e.g.~for $V_i>0$ the current flows into the $i$th load, so $I_i<0$. 
Then $J_{\rm total}=-P_{\rm total}>0$, so the quantum system absorbs the heat turned into
electrical power.

\section{Second law of thermodynamics}

\subsection{System coupled to two reservoirs}
Here I reproduce the calculation in Ref.~[\onlinecite{Bruneau2012}] 
whose outcome is the second law in two-reservoir systems.  
The Hartree-type interaction effects 
can be included without any changes to their method.
For two reservoirs, the rate of change of entropy is $\dot S= -(J_1/T_1 +J_2/T_2)$, 
one can substitute in Eq.~(\ref{Eq:J-initial}) and explicitly write out the sum over $j=1,2$.
There is no Andreev scattering so 
${\cal T}_{ij}^{\mu\nu}\propto \delta_{\mu\nu}$, 
hence when there are only two-reservoirs 
Eq.~(\ref{Eq:transmission-sum}) gives us the 
following two-terminal Onsager relation, 
$N_1-{\cal T}_{11}^{\mu\mu}=  {\cal T}_{21}^{\mu\mu} = N_2-{\cal T}_{22}^{\mu\mu}= {\cal T}_{12}^{\mu\mu}$.
As a result,
\begin{eqnarray}
{\dot{S}_{\rm total} \over \kB} &=& 
-\sum_{\mu} \int_0^\infty {{\rm d}\eps \over h}\, \big[\xi_{1\mu}-\xi_{2\mu}\big]
\nonumber \\
& & \qquad \ \ \times\ 
{\cal T}_{12}^{\mu \mu}(\eps, {\bf v}_{\rm env})   \big[ f(\xi_{1\mu}) - f(\xi_{2\mu}) \big], \qquad
\label{Eq:S-2lead}
\end{eqnarray}
where I define $\xi_{i\mu}= (\eps - \mu \eminus V_i)/(\kB T_i)$,
so $f_i^\mu(\eps)= f(\xi_{i\mu})$ for $f(\xi)= (1+\exp[\xi])^{-1}$.
Since $f(\xi)$ is a monotonically decaying function of $\xi$, 
the product of the two square-brackets in Eq.~(\ref{Eq:S-2lead}) cannot be positive.
Taking this together with Eq.~(\ref{Eq:positivity-of-T}), one concludes that the 
second law, $\dot S_{\rm total} \geq 0$, is satisfied for such a two-reservoir system.

\subsection{System coupled to any number of reservoirs}
In general for more than two reservoirs, 
there is no simple relationship between
${\cal T}_{ji}^{\nu\mu}$ and $T_{ij}^{\mu\nu}$.
Then the derivation of the second law is more involved.  
Here I provide a derivation that uses only the unitarity of ${\cal S}_{ij}^{\mu\nu}$,
in the form of Eqs.~(\ref{Eq:positivity-of-T},\ref{Eq:transmission-sum}).
Using Eq.~(\ref{Eq:J-initial}), I write
$\dot S_{\rm total} \ =\ -(\kB/h)\int_0^\infty  \rmd \eps \ Y (\eps) $ with
\begin{eqnarray}
Y(\eps) \ =\  
\sum_{ij\mu\nu}\ \xi_{i\mu} \ \big[N_i\delta_{ij}^{\mu\nu}-{\cal T}_{ij}^{\mu\nu}(\eps)  \big] \ f(\xi_{j\nu}). \qquad
\label{Eq:Y}
\end{eqnarray} 
where the $i$ and $j$ sums are over the $\Kn$ normal reservoirs.
As in Eq.~(\ref{Eq:S-2lead}),  
I defined  $\xi_{i\mu}= (\eps - \mu \eminus V_i)/(\kB T_i)$
and
$f(\xi) = \left(1+\exp\left[\xi \right] \right)^{-1}$.
I now prove that $Y (\eps) \leq 0$ for all $\eps$,
to show that the second law, Eq.~(\ref{Eq:2ndlaw}), is satisfied.

One can re-label the reservoirs in Eq.~(\ref{Eq:Y}), 
giving separate labels to the electron and hole state entering the system from a given reservoir.
Thus I replace the sum $\sum_{i=1}^{\Kn} \sum_{\mu=\pm 1}$ with
$\sum_{\tildei=1}^{2\Kn}$, where $\tildei$ is related to $i$ and $\mu$ as follows.
For each  $i$ and $\mu$, one calculates $\xi_{i\mu}$, and then
order the elements in ascending order of $\xi_{i\mu}$.
I then label these states with superscripts, $\tildei$, running from 1 to $2\Kn$,
such that $\xi^{(1)}<\xi^{(2)}< \cdots < \xi^{(2\Kn)}$. 
The quadruple sum over $i,j,\mu,\nu$ in Eq.~(\ref{Eq:Y})
can then be replaced with a double sum over $\tildei,\tildej$.
I then define the differences
\begin{eqnarray}
\Delta\xi^{(\tilden)} &=&\xi^{(\tilden)} -\xi^{(\tilden -1)}  \\
\Delta f^{(\tildem)} &=& f(\xi^{(\tildem)}) -f(\xi^{(\tildem -1)}), 
\end{eqnarray} 
and  so use
$\xi^{(\tildei)} =
\xi^{(0)}  + \sum_{\tilden=1}^{\tildei} \Delta\xi^{(\tilden)}$, 
and $
f(\xi^{(\tildei)} ) =
f(\xi^{(0)} )+ \sum_{\tildem=1}^{\tildej} \Delta f^{(\tildem)}$ to get
\begin{eqnarray}
\label{Eq:Y2}
Y(\eps) \! &=& \! 
\sum_{\tildei,\tildej=1}^{2\Kn}\bigg[
\ \xi^{(0)} \ \big[N_i\delta_{\tildei\tildej}-{\cal T}_{\tildei\tildej}(\eps)  \big] \ f(\xi^{(0)})
\\
& & \qquad
+\sum_{\tilden=1}^{\tildei} \  \Delta\xi^{(\tilden)}  \ 
 \big[N_i\delta_{\tildei\tildej}-{\cal T}_{\tildei\tildej}(\eps)  \big] \ f(\xi^{(0)})
\nonumber 
\\
& &  \qquad
+\sum_{\tildem=1}^{\tildej}  \xi^{(0)}  \ 
 \big[N_i\delta_{\tildei\tildej}-{\cal T}_{\tildei\tildej}(\eps)  \big]  \ 
\Delta f^{(\tildem)} 
\nonumber 
\\
& &  \qquad
+\sum_{\tilden=1}^{\tildei} \sum_{\tildem=1}^{\tildej} 
\Delta\xi^{(\tilden)}  \ 
 \big[N_i\delta_{\tildei\tildej}-{\cal T}_{\tildei\tildej}(\eps)  \big] \ 
\Delta f^{(\tildem)} \bigg].  \qquad
\nonumber 
\end{eqnarray} 
Here $\xi^{(0)}$ is arbitrary, so I take  $\xi^{(0)} < \xi^{(1)}$.
Then $\Delta\xi^{(\tilden)}\geq 0$ for all $\tilden$, and 
 $\Delta f^{(\tildem)}\leq0$ for all $\tildem$ (since 
$f(\xi )$ decays monotonically with $\xi$).
From Eq.~(\ref{Eq:transmission-sum}), one sees that each of the first three lines of Eq.~(\ref{Eq:Y2})
equals zero when summed over $\tildei,\tildej$; leaving only the term containing the sum over both
$\tilden$ and $\tildem$ to evaluate.
To do this, I write 
$\sum_{\tildei=1}^{2\Kn} \sum_{\tilden=1}^{\tildei} = \sum_{\tilden=1}^{2\Kn} \sum_{\tildei=\tilden}^{2\Kn}$
and
$\sum_{\tildej=1}^{2\Kn} \sum_{\tildem=1}^j = \sum_{\tildem=1}^{2\Kn} \sum_{\tildej=\tildem}^{2\Kn}$,
after which
\begin{eqnarray}
Y(\eps) \! &=& \! 
\sum_{\tilden,\tildem=1}^{2\Kn} \Delta\xi^{(\tilden)} \ \Delta f^{(\tildem)}
\sum_{\tildei=\tilden}^{2\Kn} \sum_{\tildej=\tildem}^{2\Kn} 
 \  \big[N_i\delta_{\tildei\tildej}-{\cal T}_{\tildei\tildej}(\eps)  \big].
\nonumber 
\end{eqnarray} 
For $\tilden>\tildem$, one can prove that the sum over $\tildej$ is positive (or zero) 
for all $\tildei$.
The proof requires noting that 
every sum over $\tildej$ contains a positive term for $\tildej=\tildei$ and negative terms for $\tildej \neq \tildei$,
and then using Eqs.~(\ref{Eq:positivity-of-T},\ref{Eq:transmission-sum}) 
to show that the sum over $\tildej$ is not negative.
This is not the case if $\tilden<\tildem$, because not all sums over $\tildej$ contain a term with $\tildej=\tildei$.
However, for $\tilden<\tildem$ one can do the $\tildei$ sum first;
using the same logic as above to find that the sum over $\tildei$ is positive (or zero) for all $\tildej$.
Thus $\sum_{\tildei=\tilden}^{2\Kn} \sum_{\tildej=\tildem}^{2\Kn} 
 \ \big[{\cal T}_{\tildei\tildej}(\eps) - N_i\delta_{\tildei\tildej} \big] \geq 0$ for all $\tilden,\tildem$.
Combining this with the fact that $\Delta\xi^{(\tilden)}\geq 0$ for all $\tilden$
and $\Delta f^{(\tildem)}\leq0$ for all $\tildem$, shows that $Y(\eps) \leq 0$,
and thus the second law, Eq.~(\ref{Eq:2ndlaw}), is obeyed.

On a side note, I want to mention that the above proof of
$\sum_{\tildei=\tilden}^{2\Kn} \sum_{\tildej=\tildem}^{2\Kn} 
 \ \big[N_i\delta_{\tildei\tildej}-{\cal T}_{\tildei\tildej}(\eps)  \big] \geq 0$ is most easily seen graphically.
One can write out the matrix 
$\big[N_i\delta_{\tildei\tildej}-{\cal T}_{\tildei\tildej}\big]$ explicitly and draw a rectangle around the
elements summed over for given $\tilden,\tildem$.  
For $\tilden>\tildem$, every {\it row} in that rectangle contains 
a diagonal ($\tildei=\tildej$) element, so Eqs.~(\ref{Eq:positivity-of-T},\ref{Eq:transmission-sum}) show that 
summing the elements in each {\it row} gives a positive (or zero) result.  
In contrast, for $\tilden<\tildem$,
every {\it column} of the rectangle contains a diagonal ($i=j$) element, thus summing over the elements in each {\it column} gives a positive (or zero) result.

\subsection{Emission of Joule heat}
It is common that
all reservoirs have the same temperature; $T_i=T_0$ for all $i$.
Under this circumstance, the above proof of Eq.~(\ref{Eq:2ndlaw}), 
proves that the quantum system must emit Joule heat, so $J_{\rm total} \leq 0$.
 

\section{Quantum bound on heat flows}

Eq.~(\ref{Eq:J-initial}) has an upper-bound 
on the amount of heat that a thermoelectric quantum system can extract 
from any given reservoir, Eq.~(\ref{Eq:quantumbound}). 
This is simply because no system can extract more heat from a reservoir 
than if the system is emptying every full state that comes to it at an energy above the reservoir's 
chemical potential, 
{\it and} filling every empty electron state that comes to it at an energy 
below the reservoir's chemical potential \cite{Pendry1983}.
If we can get $\eps=0$ to coincide with the $i$th reservoir's chemical potential, then the proof is easy (see how the below proof simplifies for $V_i=0$).  However,
this may not be possible when there is a superconductor present, 
or if one is interested in heat flows in multiple leads with 
different chemical potentials.  Thus below, I provide the proof for arbitrary $V_i$.

To proceed, I split Eq.~(\ref{Eq:J-initial}) into two parts,
with $J_i^{\rm (a)}$ containing only the parts of the integrals 
with $(\eps -\mu \eminus V_i)>0$, 
while $J_i^{\rm (b)}$ contains the rest.
Below I assume positive $\eminus V_i$,
the proof for negative $\eminus V_i$ follows the same logic
(under the interchange e $\leftrightarrow$ h).
For positive  $\eminus V_i$, one has
$J_i=J_i^{\rm (a)}+J_i^{\rm (b)}$ with
\begin{eqnarray}
& & \hskip -8mm  
J_i^{\rm (a)} (\eminus V_i>0) 
\nonumber \\
&=&  \! \!
\int_{\eminus V_i}^\infty {{\rm d}\eps \over h}\, (\eps- \eminus V_i) \sum_{j,\nu} 
\big[N_i\delta_{ij}^{+1,\nu} - {\cal T}_{ij}^{+1,\nu}(\eps) \big]  \, f_j^\nu (\eps)
\nonumber \\ 
& &  \! \! + \int_{0}^\infty {{\rm d}\eps \over h}\, (\eps+ \eminus V_i) \sum_{j,\nu}
\big[N_i\delta_{ij}^{-1,\nu} - {\cal T}_{ij}^{-1, \nu}(\eps) \big]  \, f_j^\nu (\eps),
\quad \ 
\nonumber 
\end{eqnarray}
\begin{eqnarray}
& & \hskip -8mm  
J_i^{\rm (b)} (\eminus V_i>0) 
\nonumber \\
&=&   \! \!
 \int_0^{\eminus V_i} {{\rm d}\eps \over h}\, (\eps- \eminus V_i) \sum_{j,\nu} 
\big[ N_i\delta_{ij}^{+1,\nu} - {\cal T}_{ij}^{+1, \nu}(\eps) \big]  \, f_j^\nu (\eps).
\nonumber 
\end{eqnarray}
where the $\pm1$ is for e and h respectively.
Our objective is to find the upper-bound on $J_i^{\rm (a)}$
and $J_i^{\rm (b)}$.

For  $J_i^{\rm (a)} (\eminus V_i>0)$, the first term in both integrands 
(the terms of the form $(\eps \pm \eminus V_i)$) is positive,
while  Eq.~(\ref{Eq:positivity-of-T}) tells us that ${\cal T}_{ij}^{\mu\nu}>0$,
and the Fermi function satisfies $0\leq f_j^\nu(\eps)\leq 1$.
Thus $J_i^{\rm (b)} (\eminus V_i>0)$
can never  be more positive than when $f_j^\nu(\eps)=0$
for all $j\neq i$ or $\nu\neq\mu$ (where $\mu=+1$ 
in the first integral and $\mu=-1$ in the second).
The remaining terms --- those with $j=i$ and $\mu=\nu$ --- 
can never be more positive than when $ {\cal T}_{ii}^{{\rm \mu\mu}} =0$.
Making the substitution $\xi= (\eps -\mu \eminus V_i)/(\kB T_i)$
for $\mu=+1$ in the first integral and $-1$ in the second, one arrives at 
\begin{eqnarray}
J_i^{\rm (a)} (\eminus V_i>0) \, 
\leq  \,
  { N_i (\kB T_i)^2 \over h} \left[ P(0) + P\left( {\eminus V_i \over \kB T_i} \right)\right] \!  , \quad
\label{Eq:Ja-limit}
 \end{eqnarray}
having defined 
$P(x) \equiv  \int_x^\infty {\rm d} \xi \ \xi \big/ (1+\exp [\xi] )$.

For $J_i^{\rm (b)} (\eminus V_i>0)$, 
the first term in the integrand (the term of the form $(\eps - \eminus V_i)$) 
is negative.
Thus $J_i^{\rm (b)} (\eminus V_i>0)$ is most positive when
the $j,\nu$-sum is as negative as possible, and 
this is when
the Fermi functions  $f_j^\nu(\eps)=1$ for all terms with $j\neq i$ or $\nu \neq +1$.
Then using Eq.~(\ref{Eq:transmission-sum}) to write the sum 
over $(j,\nu) \neq (i,\mu)$ in terms of the $(i,\mu)$ element of the sum, 
one sees that
$J_i^{\rm (b)} \leq -
 \int_0^{\eminus V_i}  {{\rm d}\eps \over h}\, (\eps- \eminus V_i) 
\big[N_i-{\cal T}_{ii}^{+1,+1}(\eps)  \big]  \, \left[1-f_i^{+1} (\eps) \right]$, which 
is most positive when ${\cal T}_{ii}^{{\rm +1,+1}}=0$. Taking this upper bound, one
can make the substitution $\xi= (\eps - \eminus V_i)/(\kB T_i)$, 
and use the fact that $[1-f(\xi)]=f(-\xi)$, thus arriving at  
\begin{eqnarray}
J_i^{\rm (b)} (\eminus V_i>0) \, 
\leq  \,
  { N_i (\kB T_i)^2 \over h} \left[ P(0) - P\left( {\eminus V_i \over \kB T_i} \right)\right] \!  , \quad
\label{Eq:Jb-limit}
 \end{eqnarray}
Taking the sum of Eqs.~(\ref{Eq:Ja-limit},\ref{Eq:Jb-limit}), and noting that 
$P(0)= -{\rm Li}_2(-1)= \pi^2/12$,  where ${\rm Li}_2(z)$ is a dilogarithm function,
one arrives at Eq.~(\ref{Eq:quantumbound}).
Redoing the derivation for $\eminus V_i <0$, gives different results for
 $J_i^{(a)}$ and $J_i^{(b)}$, but their sum is the same as for  $\eminus V_i >0$, thus
the inequality in Eq.~(\ref{Eq:quantumbound}) holds for all $V_i$.
 
For {\it passive} cooling, the bound is a fermionic
analogue of the Stefan-Boltzmann law of 
black-body radiation \cite{footnote:universalSBlaw}.
To reach the bound, $J^{\rm qb}_i$, one couples reservoir $i$  to a zero-temperature reservoir with the same chemical-potential, through a contact containing $N_i$ modes 
\cite{Pendry1983}. 
To achieve the bound with {\it active} cooling (thermoelectrically driving heat against a thermal gradient) is more complicated. I will discuss it elsewhere 
\cite{whitney-in-prep} and show that for 
coherent two-terminal devices, it cannot exceed $\half J^{\rm qb}$.

\section{Phenomenological treatment of decoherence and relaxation}

The scattering theory used above includes electrostatic interaction
effects, but not dynamic interaction effects.  
This means that neither temperature nor bias induce the decoherence or relaxation (thermalization) of the electrons as they pass through the quantum system.  
However, one can include these effects in the model in a {\it phenomenological} manner.
The idea is to model a system with such dynamic interaction effects
as follows.

Relaxation to a thermal state (and the associated decoherence) is modeled by fictitious reservoirs \cite{voltage-probes,deJong-Beenakker,Jacquet2009,Jacquet2012}, 
whose bias and temperature are chosen so  
that the average charge and heat flows into each such reservoir are zero.
``Pure-dephasing'' (decoherence with no relaxation) \cite{deJong-Beenakker}
is modeled by a large set of
fictitious reservoirs coupled to the system via leads which only let a 
given energy pass through them, and tuning the bias 
on those reservoirs so that the charge and heat flows into all of them are zero.

A system without dynamic interactions, but with the fictitious reservoirs,  
obeys the bounds in  
Eqs.~(\ref{Eq:1stlaw}-\ref{Eq:quantumbound}).
Since the average charge and
heat currents into the fictitious reservoirs are
zero, the bounds on flows into the real reservoirs are the same as in the 
absence of the fictitious reservoirs.
Thus it seems reasonable to conclude that Eqs.~(\ref{Eq:1stlaw}-\ref{Eq:quantumbound}),
are obeyed regardless of 
whether the electrons decohere and relax within the quantum system 
(due to electron-electron interactions) or not.

\section{Limits on quantum heat-engines}
\label{Sect:heat-engine}
Consider a two terminal thermoelectric device extracting electrical power 
$P_{\rm total}$  
from a heat flow between hot  (temperature $T_{\rm H}$)  and cold  (temperature $T_{\rm C}$) reservoirs.
Then, $J_{\rm H} >0$ and $J_{\rm C}<0$, while
the power generated is $P_{\rm gen}= -P_{\rm total} >0$.
The first and second laws, Eqs.~(\ref{Eq:1stlaw},\ref{Eq:2ndlaw}),
combine to give Carnot's result that
\begin{eqnarray}
 P_{\rm gen} \leq J_{\rm H} \left( 1 - {T_{\rm C}\big/T_{\rm H}} \right). 
\label{Eq:fridge-thermodyn-bound}
\end{eqnarray}
This places a bound on the efficiency of power extraction, $\eta=P_{\rm gen}/J_{\rm H}$, 
but not on the power itself. 
Yet, taking  Eq.~(\ref{Eq:fridge-thermodyn-bound}) with the quantum bound on $J_{\rm H}$ in Eq.~(\ref{Eq:quantumbound}), one gets a bound on the power itself,
\begin{eqnarray}
 P_{\rm gen} \leq { N_{\rm H} \pi^2 \kB^2 T_{\rm H} ( T_{\rm H} - T_{\rm C} ) \over 6h}.
\label{Eq:heat-engine-bound}
\end{eqnarray}
For a heat-engine between a hot-source at $T_{\rm H}=100^\circ$C$=373$K,
and a cold source at $T_{\rm C}=20^\circ$C$=293$K,
the maximum power per transverse mode is 15nW.

\section{Limits on quantum refrigerators}
\label{Sect:refrigerator}

Consider a two terminal device extracting heat from a cold reservoir 
(to keep it cold or cool it further), when the ambient temperature is $T_0$ and the cold reservoir's 
temperature is $T_{\rm C} <T_0$.   Power is supplied (positive $P_{\rm total}$)
to extract heat from the cold reservoir (positive $J_{\rm C}$).
Combining the first and second laws, Eqs.~(\ref{Eq:1stlaw},\ref{Eq:2ndlaw}),
one gets the Carnot bound on refrigeration.
Eq.~(\ref{Eq:quantumbound}) gives a second quantum bound.
Together they read,
\begin{eqnarray}
J_{\rm C} \leq \ \left\{ \begin{array}{lcl} 
 {\displaystyle{ P_{\rm total}  \  {T_{\rm C} \big/ (T_{\rm 0} -T_{\rm C})}  }}, & 
 \ & \hbox{[thermodyn.]}
\\ 
{\displaystyle{ {N_{\rm C} \pi^2 (\kB T_{\rm C})^2 \big/ 6h}  }}.& &
\hbox{[quantum]} \phantom{\Big|}
\end{array} \right.
\end{eqnarray} 
The thermodynamic bound places no limit
on the heat that one can extract from the cold reservoir ---
one can always extract heat if one is able to provide enough electrical power.  
In stark contrast the quantum bound 
places an absolute limit on the heat extracted.

Imagine a kitchen refrigerator consisting of thermoelectric devices, 
running at 100W, and 
cooling from room-temperature $T_0 = 293$K to $T_{\rm C}=0^\circ$C$=273$K,
then
\begin{eqnarray}
J_{\rm C} \leq \ \left\{ \begin{array}{lcl} 
 1300 \,{\rm W},  & \ \ & \hbox{[thermodyn.] } \phantom{\big|}
\\ 
N_{\rm C} \times 36 \,{\rm nW}. & & \hbox{[quantum]} \phantom{\Big|}
\end{array} \right.
\end{eqnarray}
The quantum bound is stronger than the thermodynamic one 
if $N_{\rm C} \lesssim  10^{10}$.
A typical thermoelectric semiconductor 
(Fermi wavelength $\sim100$nm) with a cross-section of 1cm$\times$1cm, has
 $N_{\rm C} \sim 10^{10}$ ---
the quantum bound dominates for cross-sections smaller than this.  

Consider a point-contact such as in Ref.~[\onlinecite{Kouwenhoven1989}],
where 
typical powers are pico-Watts (pW).
Suppose such a device is used for cooling as proposed
in Ref.~[\onlinecite{2012w-pointcont}], to refrigerate a micron-sized island 
from the cryostat temperature $T_0 \sim1$K
down to  $T_{\rm C}\sim 0.1$K.
Then
\begin{eqnarray}
J_{\rm C} \leq \ \left\{ \begin{array}{lcl} 
0.1 \,{\rm pW},  & \ \ & \hbox{[thermodyn.]}  \phantom{\big|}
 \\
N_{\rm C} \times 0.005 \,{\rm pW}. & & \hbox{[quantum]}  \phantom{\Big|}
\end{array} \right.
\end{eqnarray}
For a single-mode point contact, the quantum bond is more important. However the thermodynamic one dominates for 20 or more such point-contacts
in parallel thermally (as in Fig. 2b of Ref.~[\onlinecite{2012w-pointcont}]).

\section{Nernst's unattainability principle}

Nernst's unattainability principle, is one of the 
two statements that are together referred to as the third law of thermodynamics.  
It says that one can never take a finite temperature reservoir down to
zero temperature in a finite time.
(The other part of the third law is the claim that entropy vanishes at zero temperature.) 
The textbook result for the heat capacity of a reservoir of free electrons
is $C_i  \equiv (\rmd Q_i/ \rmd T_i) \propto T_i$.   
If the quantum system is extracting heat out over a long time, then
the fastest rate of heat removal is given by quantum bound, 
$J^{\rm qb}_i \propto T_i^2$.
Assuming this heat flow is weak enough that the reservoir remains in equilibrium,
we have $-J^{\rm qb}_i = (\rmd Q_i/ \rmd t) = C_i (\rmd T_i/ \rmd t)$.
Thus 
\begin{eqnarray}
{\rmd T_i \over \rmd t} \propto -T_i^{\zeta}, \qquad \hbox{ with } \zeta=1.
\label{Eq:unattain}
\end{eqnarray}
For arbitrary $\zeta$, 
one sees that $T_i=0$ is reached in a finite time if $\zeta <1$, 
but not if $\zeta>1$.   
Eq.~(\ref{Eq:unattain}) has the critical value $\zeta=1$, then the temperature decays exponentially with time $t$;
thus it gets exponentially close to $T_i=0$, but never quite reaches it.
The point-contact in Ref.~[\onlinecite{2012w-pointcont}] at large bias has  
$\zeta=1$.  Scattering matrices with smoother $\eps$-dependences give larger $\zeta$, so the decay goes like $t^{-1/(\zeta-1)}$ for large $t$. 
The systems considered here never have $\zeta < 1$
and so never violate the unattainability principle
in the manner proposed in Ref.~[\onlinecite{Kolar2012}].

\section{Concluding remarks}
\label{Sect:concluding}

This work is on the fully nonlinear zero-frequency (DC) heat and charge response to time-independent potentials.
It applies for any quantum device
where single-electron charging effects are not significant
(i.e.\ the single-electron charging energy is much less than the temperature or level-broadening 
due to the coupling to reservoirs).
We do not consider heat-engines and refrigerators that rely on a  classical or quantum pumping cycle,
although work in this direction is desirable.
One should also remember that quantum transport is inherently noisy, so the
 energy and entropy in the quantum system fluctuate randomly on short time scales.
This noise has little effect at low-frequency, as it averages out over long time, 
but it will affect finite-frequency responses.

Finally, it is worth recalling that all the results presented here 
rely on only two ingredients;
(i) the Schrodinger's equation formulated as a scattering theory,
and (ii) simple {\it equilibrium} statistical mechanics in the reservoirs.
It is remarkable that the thermodynamic laws
for {\it non-equilibrium} transport (far outside the linear-response regime) 
emerge so naturally from a simple combination of these two ingredients.




\begin{thebibliography}{1}
\bibitem{footnote:DCresponse}
In this work ``DC'' is used to mean the zero-frequency 
heat and charge currents due to time-independent biases and 
temperature differences. In Sect.~\ref{Sect:concluding}, 
I briefly mention the limitations of studying only this DC response.
\bibitem{footnote:universalSBlaw}
Stefan-Boltzmann law for
black-body radiation (or relativistic fermions\cite{Neutrinos}  at $\kB T \gg E_{\rm F}$) has $J \propto T^4$.
All such laws take the form
$J \propto N \ (k_{\rm B}T)^2$, with $N\sim A/\lambda_{\rm deBr}^2$
for aperture cross-section $A$.
The de\,Broglie wavelength $\lambda_{\rm deBr} \sim h\big/\sqrt{2mE_{\rm F}}$ for 
non-relativistic fermions with $E_{\rm F} \gg \kB T$, 
while $\lambda_{\rm deBr} \sim hc\big/(\kB T)$
for relativistic fermions with $\kB T \gg E_{\rm F}$
or photons.
\bibitem{footnote:finite-SC-gap}
I assume the sc reservoir is a superconducting condensate 
without quasi-particles, i.e. sc gap  
much larger than temperatures and biases. 
Quasiparticles in the superconductor can be treated 
as a separate electronic reservoir.
\end{thebibliography}
\end{document}